\documentclass[prl, nofootinbib, twocolumn,floatfix,preprintnumbers,superscriptaddress,bibliography]{revtex4-1}
\usepackage[colorlinks, linkcolor=red, anchorcolor=blue, citecolor=blue, colorlinks=true, urlcolor=blue]{hyperref}
\usepackage{amssymb}
\usepackage{amsmath}
\usepackage{amsthm}
\usepackage{lipsum}
\usepackage{graphicx}
\usepackage{indentfirst}
\usepackage{subfigure}
\usepackage{ulem}
\usepackage{bm}
\usepackage{mathrsfs}
\usepackage{pifont}
\usepackage{threeparttable}
\usepackage{bbding}
\usepackage{gensymb}
\usepackage{breakurl}
\usepackage{epstopdf}
\usepackage[figuresright]{rotating}
\usepackage{fontawesome}

\begin{document}

\title{Upper limit on the axion-photon coupling from Markarian\,421}

\author{Hai-Jun Li} 
\email{lihaijun@itp.ac.cn}
\affiliation{Key Laboratory of Theoretical Physics, Institute of Theoretical Physics, Chinese Academy of Sciences, Beijing 100190, China}

\author{Wei Chao} 
\email{chaowei@bnu.edu.cn}
\affiliation{Center for Advanced Quantum Studies, Department of Physics, Beijing Normal University, Beijing 100875, China}

\author{Yu-Feng Zhou}
\email{yfzhou@itp.ac.cn}
\affiliation{Key Laboratory of Theoretical Physics, Institute of Theoretical Physics, Chinese Academy of Sciences, Beijing 100190, China}
\affiliation{School of Physical Sciences, University of Chinese Academy of Sciences, Beijing 100049, China}
\affiliation{School of Fundamental Physics and Mathematical Sciences, Hangzhou Institute for Advanced Study, UCAS, Hangzhou 310024, China}
\affiliation{International Centre for Theoretical Physics Asia-Pacific, Beijing/Hangzhou, China}

\preprint{ITP-24-070, BNU-24-039}

\date{\today}

\begin{abstract}

Markarian\,421 is a well-known nearby BL Lac blazar at the redshift $z=0.031$.
Many previous works were investigated to constrain the axion-photon coupling from its TeV gamma-ray observations, showing the upper limit on the coupling constant $g_{a\gamma} \lesssim 2.0\times 10^{-11} \rm \, GeV^{-1}$ for the axion mass $[5.0\times10^{-10} \, {\rm eV} \lesssim m_a \lesssim 5.0\times10^{-7} \, {\rm eV}]$.
While in this work, we obtain a more stringent upper limit on the axion-photon coupling from the 1038 days gamma-ray observations of the blazar Markarian\,421.
The long-term gamma-ray spectra are measured by the collaborations Large Area Telescope on board NASA's Fermi Gamma-ray Space Telescope (Fermi-LAT) and High Altitude Water Cherenkov (HAWC) Gamma-Ray Observatory from 2015 June to 2018 July.
We show the best-fit spectral energy distributions (SEDs) of Markarian\,421 under the null and axion hypotheses.
Then we set the axion-photon limit in the $\{m_a, \, g_{a\gamma}\}$ plane.
The 99\% $\rm C.L.$ upper limit set by Markarian\,421 is $g_{a\gamma} \lesssim 4.0\times 10^{-12} \rm \, GeV^{-1}$ for the axion mass $[1.0\times10^{-9} \, {\rm eV} \lesssim m_a \lesssim 1.0\times10^{-8} \, {\rm eV}]$.
It is the most stringent upper limit in this axion mass region. 


\end{abstract}
\maketitle


\medskip\noindent{\bf Introduction.}---%
The pseudo-Nambu-Goldstone bosons (pNGBs), axions, are motivated candidates for physics beyond the Standard Model (SM).
The QCD axion \cite{Peccei:1977hh, Peccei:1977ur, Weinberg:1977ma, Wilczek:1977pj} was originally postulated to solve the strong CP problem, meanwhile, it is the cold dark matter (DM) candidate \cite{Preskill:1982cy, Abbott:1982af, Dine:1982ah}.
Additionally, the axionlike particle (ALP) \cite{Arvanitaki:2009fg, Svrcek:2006yi} can also account for the DM, but is not associated to the solution of the strong CP problem.
For simplicity, in the following the ``axion" stands for the ALP.
The axion-photon interaction has a two-photon vertex
\begin{eqnarray}
\mathcal{L}_{\rm }\supset-\dfrac{1}{4}g_{a\gamma} a F_{\mu\nu}\tilde{F}^{\mu\nu}\, ,
\end{eqnarray}
where $a$ is the axion field, $g_{a\gamma}$ is the axion-photon coupling constant, $F_{\mu\nu}$ is the electromagnetic field tensor and its dual tensor $\tilde{F}^{\mu\nu}=\frac{1}{2}\epsilon^{\mu\nu\rho\sigma}F_{\rho\sigma}$. 
Considering the interaction between the axion and very high energy (VHE) photon in the astrophysical magnetic fields, it can lead to some detectable effects, such as the reduced TeV opacity of the Universe \cite{Mirizzi:2007hr, Hooper:2007bq, Simet:2007sa, Fairbairn:2009zi}.
The VHE TeV gamma-ray emissions from the extragalactic space, such as the active galactic nuclei (AGN), are mainly affected by the effect of extragalactic background light (EBL) absorption through the electron-positron pair production process $\gamma_{\rm TeV} + \gamma_{\rm EBL} \to e^+ + e^-$.
If considering the axion-photon conversion and further back-conversion in several astrophysical magnetic fields, the EBL absorption effect can be reduced and the Universe would appear to be more transparent than expected based on the pure EBL absorption \cite{HESS:2007xak, MAGIC:2008sib}.
Meanwhile, it provides a natural mechanism to constrain the axion properties \cite{Kohri:2017ljt, Zhang:2018wpc, Libanov:2019fzq, Long:2019nrz, Guo:2020kiq, Li:2022pqa, Cheng:2020bhr, Jacobsen:2022swa, Mastrototaro:2022kpt, LHAASO:2023lkv, Pant:2023lnz, Li:2023qyr}, see also Ref.~\cite{Galanti:2022ijh} for a recent review.

In this letter, we obtain a stringent upper limit on the axion-photon coupling from the TeV blazar Markarian\,421 with the 1038 days VHE gamma-ray observations.
The source Markarian\,421 ($\rm R.A.=11^h04^m27.31^s$, $\rm Dec.=+38^\circ12'31.8''$, J2000) is a well-known nearby blazar at the redshift $z=0.031$ \cite{1991rc3..book.....D}, it is also known as other names, Mrk\,421, TeV\,J1104+382, 1ES\,1101+384, and PG\,1101+385. 
Markarian\,421 belongs to a subclass of AGNs, the high-frequency peaked BL Lac (HBL) or the high-synchrotron peaked BL Lac (HSP) object, which are classified by the weak or missing broad emission lines in the optical spectra \cite{Urry:1995mg, Abdo:2009iq}. 
It was first discovered with the VHE emission by the Whipple 10-m Observatory in 1992 \cite{Punch:1992xw}.
Many previous works were investigated to constrain the axion-photon coupling from the gamma-ray of Markarian\,421 \cite{Li:2020pcn, Gao:2023dvn}.
Recently, its long-term gamma-ray spectra are measured by the collaborations, the Large Area Telescope on board NASA's Fermi Gamma-ray Space Telescope (Fermi-LAT) and the High Altitude Water Cherenkov (HAWC) Gamma-Ray Observatory, with the 1038 days of exposure from 2015 June to 2018 July \cite{HAWC:2021obx}.
Using this long-term gamma-ray flux, we consider the axion-photon conversion effect on the spectral energy distributions (SEDs) and set the limit on the axion-photon coupling.

\begin{figure}[t]
\centering
\includegraphics[width=0.49\textwidth]{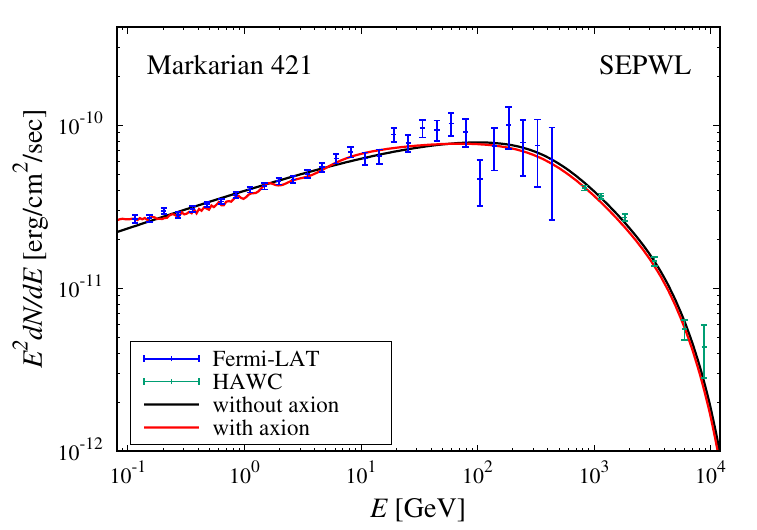}
\caption{The best-fit VHE gamma-ray SEDs of Markarian\,421 under the null and axion hypotheses.
The black and red lines represent the best-fit SEDs without/with the axion-photon conversion, respectively.
The blue and green experimental data are taken from Fermi-LAT and HAWC \cite{HAWC:2021obx}, respectively.}
\label{fig_dNdE_mrk421}
\end{figure} 

\medskip\noindent{\bf Null hypothesis.}---%
We first show the long-term VHE gamma-ray observations of Markarian\,421 under the null hypothesis.
The experimental data of Markarian\,421 measured by Fermi-LAT and HAWC are shown in Fig.~\ref{fig_dNdE_mrk421} with the blue and green points, respectively.
To fit the gamma-ray spectrum under the null hypothesis, the intrinsic spectral model is selected as the power law with a super-exponential cut-off (SEPWL)
\begin{eqnarray}
\Phi_{\rm int}(E)=N_0\left(E/E_0\right)^{-\Gamma}\exp\left(-\left(E/E_c\right)^d\right)\, , 
\end{eqnarray}
where $E$ represents the VHE photon energy, $N_0$ is a normalized constant, $\Gamma$ is the spectral index, $E_c$ and $d$ are free parameters.
Since the parameter $E_0$ can be degenerated into $N_0$, here we fix it with a typical value $E_0=1\, \rm GeV$.
As mentioned before, the main effect on VHE photon in the extragalactic space is the EBL photon absorption effect with an absorption factor $e^{-\tau}$, where $\tau$ is the optical depth, and the EBL spectral model is taken as Franceschini-08 \cite{Franceschini:2008tp}.
Then we have the chi-square value under the null hypothesis 
\begin{eqnarray}
\chi_{\rm null}^2 = \sum_{i=1}^{N}\left(\left(e^{-\tau}\Phi_{\rm int}(E_i) - \psi(E_i)\right)/\delta(E_i)\right)^2\, ,
\end{eqnarray}
where $N=36$ (Fermi-LAT: 30, and HAWC: 6) represents the gamma-ray spectral point number, $\psi$ and $\delta$ represent the detected flux and its uncertainty, respectively.
Note that here the gamma-ray intrinsic spectral model $\Phi_{\rm int}(E)$ is selected with the minimum best-fit reduced $\chi_{\rm null}^2$ from several common spectral models \cite{Li:2021zms}.
We show the best-fit gamma-ray SED of Markarian\,421 in Fig.~\ref{fig_dNdE_mrk421}.
The black line represents the best-fit SED under the null hypothesis with $\chi_{\rm null}^2=44.24$ and $\chi_{\rm null}^2/{\rm d.o.f.}=1.38$, where $\rm d.o.f.$ is the degree of freedom.

\begin{figure}[t]
\centering
\includegraphics[width=0.49\textwidth]{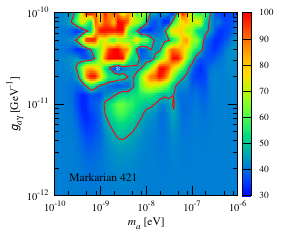}
\caption{The best-fit chi-square $\chi_{\rm axion}^2$ distribution in the $\{m_a, \, g_{a\gamma}\}$ plane for Markarian\,421. 
The red contours represent the 99\% $\rm C.L.$ exclusion regions.
The label ``$*$" corresponds to the minimum best-fit chi-square $\chi^2_{\rm min}=31.47$ at $\{m_a \simeq 2.5\times10^{-9} \, {\rm eV}, \, g_{a\gamma} \simeq 2.5\times 10^{-11} \rm \, GeV^{-1}\}$.}
\label{fig_contour_mrk421}
\end{figure}

\medskip\noindent{\bf Axion hypothesis.}---%
Next, we consider the effects of axion-photon conversion on VHE gamma-ray propagations from the source Markarian\,421 to the Earth.
In the transverse homogeneous magnetic field, the axion-photon conversion probability can be simply characterized as 
\begin{eqnarray}
\mathcal{P}_{a\gamma}=\left(g_{a\gamma}B_T L_{\rm osc}/2\pi\right)^2 \sin^2\left(\pi x_3/L_{\rm osc}\right)\, ,
\end{eqnarray}
where $B_T$ is the transverse magnetic field, $L_{\rm osc}$ is the oscillation length, and $x_3$ is the direction of the axion-photon propagation.
While in the inhomogeneous astrophysical magnetic field, the magnetic field can be modeled as the domain-like structure and each domain can be regarded as homogeneous.
In this case, the final photon-axion-photon conversion probability (or the final photon survival probability) can be described in the density matrix formal \cite{DeAngelis:2011id}
\begin{eqnarray}
\mathcal{P}_{\gamma\gamma}={\rm Tr}\left(\left(\rho_{11}+\rho_{22}\right)\mathcal{T}(s)\rho(0)\mathcal{T}^\dagger(s)\right)\, ,
\end{eqnarray}
with the whole transfer matrix
\begin{eqnarray}
\mathcal{T}(s)=\prod\mathcal{T}_i(s_i)\, ,
\end{eqnarray}
and the initial and final axion-photon density matrices
\begin{eqnarray}
\rho(0)=\dfrac{1}{2}{\rm diag}(1,1,0)\, , \quad \rho(s)=\mathcal{T}(s)\rho(0)\mathcal{T}^\dagger(s)\, ,
\end{eqnarray}
where $s$ represents the axion-photon propagation distance, and $\rho_{ii}={\rm diag}\left(\delta_{i1},\delta_{i2},0\right)$.

\begin{figure*}[t]
\centering
\includegraphics[width=0.78\textwidth]{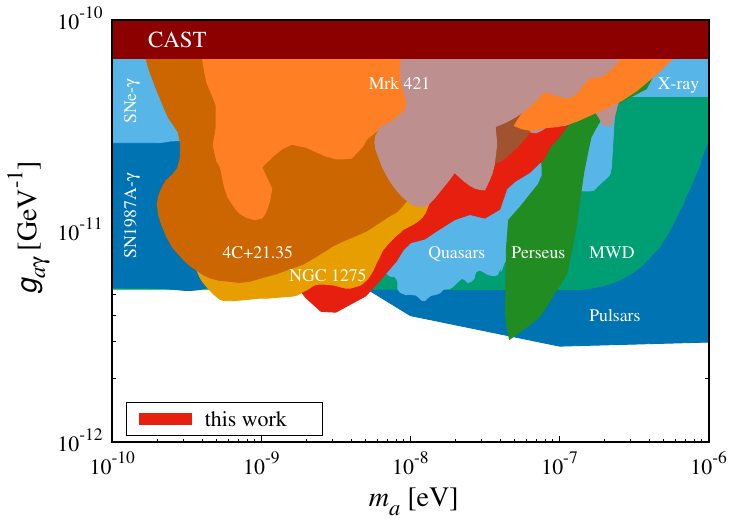}
\caption{The axion-photon limits in the $\{m_a, \, g_{a\gamma}\}$ plane.
The red contour represents the 99\% $\rm C.L.$ limit set by Markarian\,421 with the 1038 days VHE gamma-ray observations measured by Fermi-LAT and HAWC.
Other limits in the plane are taken from Ref.~\cite{ciaran_o_hare_2020_3932430}, with CAST \cite{Anastassopoulos:2017ftl}, SNe-$\gamma$ \cite{Meyer:2020vzy}, Mrk\,421 \cite{Li:2020pcn, Li:2021gxs}, X-ray \cite{Dessert:2021bkv}, SN1987A-$\gamma$ \cite{Payez:2014xsa}, 4C+21.35 \cite{Li:2022jgi}, NGC\,1275 \cite{TheFermi-LAT:2016zue}, Quasars \cite{Davies:2022wvj}, Perseus \cite{MAGIC:2024arq}, MWD \cite{Dessert:2022yqq}, and Pulsars \cite{Noordhuis:2022ljw}.}
\label{fig_compare_mrk421}
\end{figure*}

In this work, the axion-photon propagation process from Markarian\,421 to the Earth can be divided into three parts, the blazar jet magnetic field, the extragalactic space, and the magnetic field in the Milky Way.
We first discuss the axion-photon conversion in the blazar jet magnetic field of Markarian\,421, which can be described by the poloidal and toroidal components.
Here we consider a transverse magnetic field model $B(r) = B_0(r/r_{\rm VHE})^{-1}$ with an electron density model $n_{\rm el}(r) = n_0(r/r_{\rm VHE})^{-2}$, where $r_{\rm VHE}\simeq R_{\rm VHE}/\theta_{\rm jet}$ corresponds to the distance from the source central black hole to the VHE emission region, $R_{\rm VHE}$ represents the radius of the VHE emission region, $\theta_{\rm jet}$ represents the angle between the jet axis and the line of sight, $B_0$ and $n_0$ correspond to the core magnetic field and electron density at $r_{\rm VHE}$, respectively.
We also consider the energy transformation between the laboratory and co-moving frames, $E_L$ and $E_j$, with the Doppler factor $\delta_{\rm D}=E_L/E_j$.
For Markarian\,421, we take $B_0=24\pm6 \, \rm mG$, $n_0\simeq1\times10^3\, \rm cm^{-3}$, $R_{\rm VHE}=0.5\times10^{17}\, \rm cm$, $\theta_{\rm jet}=2.0^\circ$, $r_{\rm VHE}\simeq14.3\times10^{17}\, \rm cm$, and $\delta_{\rm D}=25\pm1$ \cite{HAWC:2021obx}.
Note that for the jet region at $r > 1\rm\, kpc$, the magnetic field is weak and can be taken as zero.
Then for the host galaxy region of Markarian\,421, we neglect the axion-photon conversion in this part. 
Secondly, in the extragalactic space, we just consider the EBL absorption effect on VHE photon through the pair-production process.
Since the magnetic field strength in this region is small $\sim\mathcal{O}(1)\, \rm nG$ \cite{Ade:2015cva, Pshirkov:2015tua}, the axion-photon conversion effect will be weak and can be neglected.
Thirdly, in the magnetic field of the Milky Way, we consider the axion-photon conversion again.
Here the Galactic magnetic field is simulated with the disk and halo components, and also the ``X-field" component at the Galactic center \cite{Jansson:2012pc, Jansson:2012rt, Planck:2016gdp}.

\medskip\noindent{\bf Results.}---%
Using the final photon survival probability $\mathcal{P}_{\gamma\gamma}$ discussed above, now we can obtain the chi-square value under the axion hypothesis
\begin{eqnarray}
\chi_{\rm axion}^2 = \sum_{i=1}^{N}\left(\left(\mathcal{P}_{\gamma\gamma}\Phi_{\rm int}(E_i) - \psi(E_i)\right)/\delta(E_i)\right)^2\, .
\end{eqnarray}
For one axion $\{m_a, \, g_{a\gamma}\}$ parameter set, we can derive the best-fit chi-square $\chi_{\rm axion}^2$, and also the best-fit chi-square distribution in the parameter plane.
In Fig.~\ref{fig_contour_mrk421}, we show the best-fit chi-square distribution for Markarian\,421 in the $\{m_a, \, g_{a\gamma}\}$ plane, where the label ``$*$" represents the minimum best-fit chi-square $\chi^2_{\rm min}=31.47$ at $\{m_a \simeq 2.5\times10^{-9} \, {\rm eV}, \, g_{a\gamma} \simeq 2.5\times 10^{-11} \rm \, GeV^{-1}\}$.
Meanwhile, we also show the best-fit gamma-ray SED corresponding to $\chi^2_{\rm min}$ in Fig.~\ref{fig_dNdE_mrk421} with the red line.
Compared with the null hypothesis $\chi_{\rm null}^2=44.24$ (black line), the minimum best-fit chi-square under the axion hypothesis can be significantly depressed.
Using the best-fit chi-square distribution, here we set the 99\% confidence level ($\rm C.L.$) limit on the axion-photon coupling.

In order to obtain the value of the threshold chi-square $\chi^2_{99\%}$, 400 sets of the VHE gamma-ray observations of Markarian\,421 in the pseudo-experiments by Gaussian samplings are simulated to derive the distribution of the test statistic (TS)
\begin{eqnarray}
{\rm TS}={\widehat{\chi}_{\rm null}}^2 - {\widehat{\chi}_{\rm axion}}^2\, ,
\end{eqnarray}
where ${\widehat{\chi}_{\rm null}}^2$ and ${\widehat{\chi}_{\rm axion}}^2$ are the best-fit chi-squares of the null and axion hypotheses in the Monte Carlo simulations, respectively.
It obeys the non-central chi-square distribution with the non-centrality $\lambda$ and the effective $\rm d.o.f.$, and we use it to derive the 99\% $\rm C.L.$ chi-square difference $\Delta\chi^2_{99\%}$.
Then the 99\% $\rm C.L.$ threshold chi-square is given by 
\begin{eqnarray}
\chi^2_{99\%}=\chi^2_{\rm min}+\Delta\chi^2_{99\%}\, .
\end{eqnarray}
In our simulations, we have the non-centrality $\lambda=0.01$ and the effective $\rm d.o.f. = 5.29$, indicating the threshold chi-square $\chi^2_{99\%}=46.98$.
Then we show in Fig.~\ref{fig_contour_mrk421} the 99\% $\rm C.L.$ limit with the red contours, the region in the contour is excluded.
Finally, we have the 99\% $\rm C.L.$ upper limit on the axion-photon coupling set by Markarian\,421 with the 1038 days VHE gamma-ray observations measured by Fermi-LAT and HAWC, which is constrained to the coupling constant 
\begin{eqnarray}
g_{a\gamma} \lesssim 4.0\times 10^{-12} \rm \, GeV^{-1} \, (99\% \, \rm C.L.) \, ,
\end{eqnarray}
for the axion mass 
\begin{eqnarray}
[1.0\times10^{-9} \, {\rm eV} \lesssim m_a \lesssim 1.0\times10^{-8} \, {\rm eV}]\, .
\end{eqnarray}
Compared with the latest axion-photon limits in the $\{m_a, \, g_{a\gamma}\}$ plane, see also Fig.~\ref{fig_compare_mrk421} with the red contour, it is the most stringent upper limit in this mass region.
In addition, we also have an exclusion region in the axion mass $[2.0\times10^{-10} \, {\rm eV} \lesssim m_a \lesssim 1.0\times10^{-8} \, {\rm eV}]$.

\begin{figure}[t]
\centering
\includegraphics[width=0.485\textwidth]{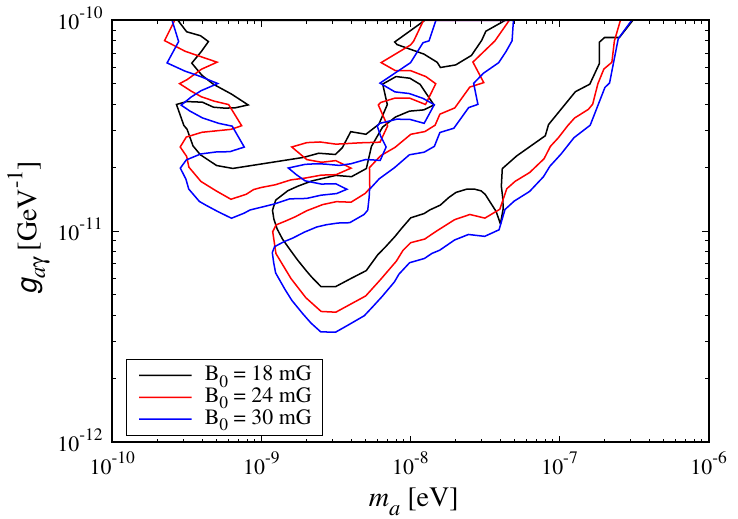}
\caption{The 99\% $\rm C.L.$ axion-photon limits in the $\{m_a, \, g_{a\gamma}\}$ plane for Markarian\,421 with $B_0=18 \, \rm mG$, $24 \, \rm mG$, and $30 \, \rm mG$.}
\label{fig_compare_mrk421_SM}
\end{figure}

We know that the greatest impact on the final result are parameters $B_0$ and $r_{\rm VHE}$, both larger of them will lead to more stringent limit.
Since we take the Markarian\,421 blazar jet core magnetic field at $r_{\rm VHE}$ as $B_0=24\pm6 \, \rm mG$, here we discuss the impact of the uncertainty of $B_0$. 
Compared with $B_0=24 \, \rm mG$, the 99\% $\rm C.L.$ axion-photon limits are shown in Fig.~\ref{fig_compare_mrk421_SM} with different lines.
As we expected, the larger $B_0$ leads to the more stringent limit.
The parameter $r_{\rm VHE}$ has a similar impact on the final result.
In addition, since we take $\delta_{\rm D}=25\pm1$ and it has little impact on the limit, here we do not show the impact of the uncertainty of $\delta_{\rm D}$.

\medskip\noindent{\bf Discussion.}---%
This is not the first investigation to constrain the axion-photon coupling from the TeV blazar Markarian\,421.
In Ref.~\cite{Li:2020pcn}, they first investigated the axion-photon conversion effect from this source by using the 4.5 years gamma-ray data of Fermi-LAT and Astrophysical Radiation with Ground-based Observatory at YangBaJing (ARGO-YBJ), showing the upper limit $g_{a\gamma} \lesssim 2.0\times 10^{-11} \rm \, GeV^{-1}$ for the axion mass $[5.0\times10^{-10} \, {\rm eV} \lesssim m_a \lesssim 5.0\times10^{-7} \, {\rm eV}]$.
In Ref.~\cite{Gao:2023dvn}, they presented the axion limits from Markarian\,421 by using the 1 year data of Fermi-LAT and Major Atmospheric Gamma Imaging Cherenkov Telescopes (MAGIC), showing the similar upper limits but for the axion mass $[8.0\times10^{-9} \, {\rm eV} \lesssim m_a \lesssim 2.0\times10^{-7} \, {\rm eV}]$.

While in this work, we obtain a more stringent upper limit from the VHE gamma-ray data of Fermi-LAT and HAWC, which is even the most stringent limit compared with other astrophysical results in the axion mass $[1.0\times10^{-9} \, {\rm eV} \lesssim m_a \lesssim 1.0\times10^{-8} \, {\rm eV}]$.
This is mainly because the large value of $r_{\rm VHE}$ obtained in this work.
On the other hand, compared with the experimental data of ARGO-YBJ and MAGIC, the data of HAWC shows smaller uncertainty in the high energy region.
Additionally, since the VHE gamma-ray observations of another TeV blazar Markarian\,501 (with redshift $z=0.034$) at the same time are measured by Fermi-LAT and HAWC in Ref.~\cite{HAWC:2021obx}, we also make the axion analysis for this source but no stringent limits are obtained.

\medskip\noindent{\bf Conclusion.}---%
In summary, we have obtained a stringent upper limit on the axion-photon coupling from the 1038 days gamma-ray observations of the TeV blazar Markarian\,421.
The long-term VHE gamma-ray spectra are measured by the collaborations Fermi-LAT and HAWC from 2015 June to 2018 July.
We show the best-fit SEDs of Markarian\,421 under the null and axion hypotheses.
Then we set the axion-photon limit in the $\{m_a, \, g_{a\gamma}\}$ plane.
The 99\% $\rm C.L.$ upper limit set by Markarian\,421 is $g_{a\gamma} \lesssim 4.0\times 10^{-12} \rm \, GeV^{-1}$ for the axion mass $[1.0\times10^{-9} \, {\rm eV} \lesssim m_a \lesssim 1.0\times10^{-8} \, {\rm eV}]$.
It is the most stringent upper limit in this axion mass region. 
 
\medskip\noindent{\bf Acknowledgments.}---%
The authors would like to thank Sara Couti$\rm \tilde{n}$o de Le$\rm \acute{o}$n for providing the experimental data of Fermi-LAT and HAWC.
W.C. is supported by the National Natural Science Foundation of China (Grants No.~11775025 and No.~12175027).
Y.F.Z. is supported by the National Key R\&D Program of China (Grant No.~2017YFA0402204), the CAS Project for Young Scientists in Basic Research YSBR-006, and the National Natural Science Foundation of China (Grants No.~11821505, No.~11825506, and No.~12047503).

\bibliography{references}

\clearpage
\maketitle
\onecolumngrid
\begin{center}
\textbf{\large Upper limit on the axion-photon coupling from Markarian\,421} \\ 
\vspace{0.05in}
{ \it \large Supplemental Material}\\ 
\vspace{0.05in}
{Hai-Jun Li, Wei Chao, and Yu-Feng Zhou}
\end{center}
\onecolumngrid
\setcounter{equation}{0}
\setcounter{figure}{0}
\setcounter{table}{0}
\setcounter{section}{0}
\makeatletter
\renewcommand{\theequation}{S\arabic{equation}}
\renewcommand{\thefigure}{S\arabic{figure}}

This Supplemental Material is organized as follows. 
In Sec.~\ref{sm_sec1}, we show the gamma-ray SEDs of Markarian\,421 under the null hypothesis.
In Sec.~\ref{sm_sec2}, we display the final photon survival probability.
In Sec.~\ref{sm_sec3}, we show the TS distribution.
In Sec.~\ref{sm_sec4}, we show the impact of the parameter uncertainty of $B_0$.
Then we make the axion analysis for the source Markarian\,501 in Sec.~\ref{sm_sec5}.
Finally, we attach the experimental gamma-ray data of Markarian\,421 and Markarian\,501 in Sec.~\ref{sm_sec6}.
 
\section{SEDs under the null hypothesis}
\label{sm_sec1}

In this section, we show the gamma-ray SEDs of Markarian\,421 under the null hypothesis with the different intrinsic spectral models.
It can be described by the simple and smooth concave functions with three to five free parameters.
We take the common spectral models as the power law with exponential cut-off (EPWL, with three parameters), the power law with super-exponential cut-off (SEPWL, with four parameters), the log-parabola (LP, with four parameters), and the log-parabola with exponential cut-off (ELP, with five parameters),
\begin{eqnarray}
{\rm EPWL}: \Phi_{\rm int} (E) &=& N_0\left(E/E_0\right)^{-\Gamma}\exp\left(-E/E_c\right)\, ,\\
{\rm SEPWL}: \Phi_{\rm int} (E) &=& N_0\left(E/E_0\right)^{-\Gamma}\exp\left(-\left(E/E_c\right)^d\right)\, ,\\
{\rm LP}: \Phi_{\rm int} (E) &=& N_0\left(E/E_0\right)^{-\Gamma-b \log\left(E/E_0\right)}\, ,\\
{\rm ELP}: \Phi_{\rm int} (E) &=& N_0\left(E/E_0\right)^{-\Gamma-b \log\left(E/E_0\right)}\exp\left(-E/E_c\right)\, ,
\end{eqnarray}
where $N_0$, $\Gamma$, $E_c$,  $b$, and $d$ are free parameters. 
For $E_0$, it is a free parameter in the models LP and ELP, while in EPWL and SEPWL, we fix $E_0 = 1\, \rm GeV$ as a typical value.
Now considering two values of $E_0$ and $E_1$ ($E_0\neq E_1\neq 0$) in SEPWL (EPWL), for simplicity, with
\begin{eqnarray}
\Phi_{\rm int,0} (E) \propto N_0\left(E/E_0\right)^{-\Gamma}=N_0 E_0^{\Gamma} E^{-\Gamma}\, ,
\label{Phi_0}
\end{eqnarray}
\begin{eqnarray}
\Phi_{\rm int,1} (E) \propto N_1\left(E/E_1\right)^{-\Gamma}=N_1 E_1^{\Gamma} E^{-\Gamma}=N_1(E_1/E_0)^{\Gamma} E_0^{\Gamma} E^{-\Gamma}\, , 
\label{Phi_1}
\end{eqnarray}
we find that if we have 
\begin{eqnarray}
N_1=(E_0/E_1)^{\Gamma} N_0\, ,
\label{N1}
\end{eqnarray}
Eq.~(\ref{Phi_1}) is equal to Eq.~(\ref{Phi_0}), which indicates that $E_0$ ($E_1$) can be degenerated into $N_0$ ($N_1$).
In other words, in the fitting as long as Eq.~(\ref{N1}) is satisfied, there is no difference between the choices of $E_0$ and $E_1$.
In order to reduce the number of parameters, we fix a typical value of $E_0 = 1\, \rm GeV$ in EPWL and SEPWL.
However, since the exponential parts of the models LP and ELP contain the term $\sim -b \log\left(E/E_0\right)$, the above derivation is not suitable for them and $E_0$ in this case must be treated as a free parameter.

\begin{figure*}[h]
\centering
\includegraphics[width=0.333\textwidth]{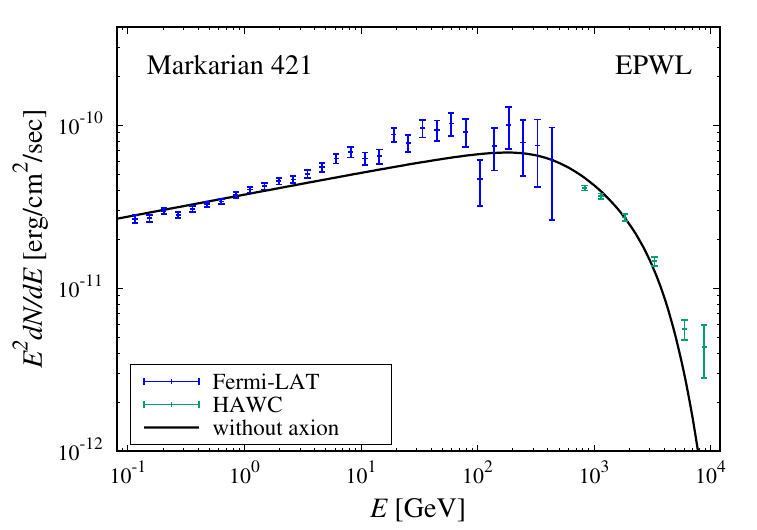}\includegraphics[width=0.333\textwidth]{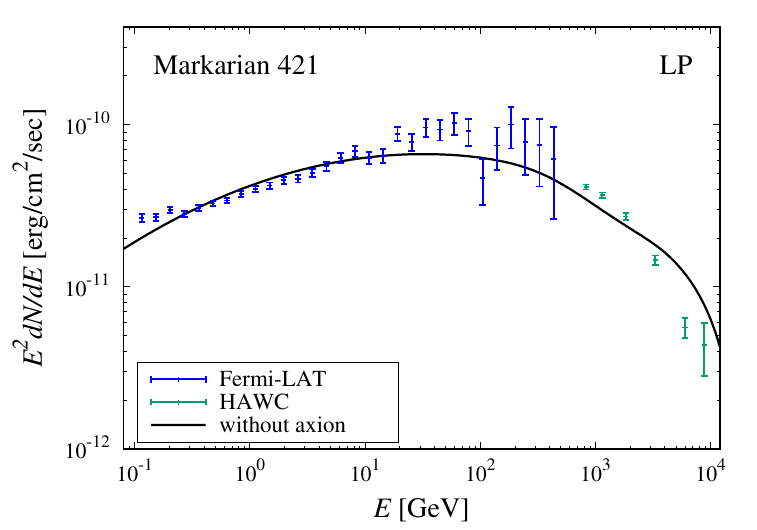}\includegraphics[width=0.333\textwidth]{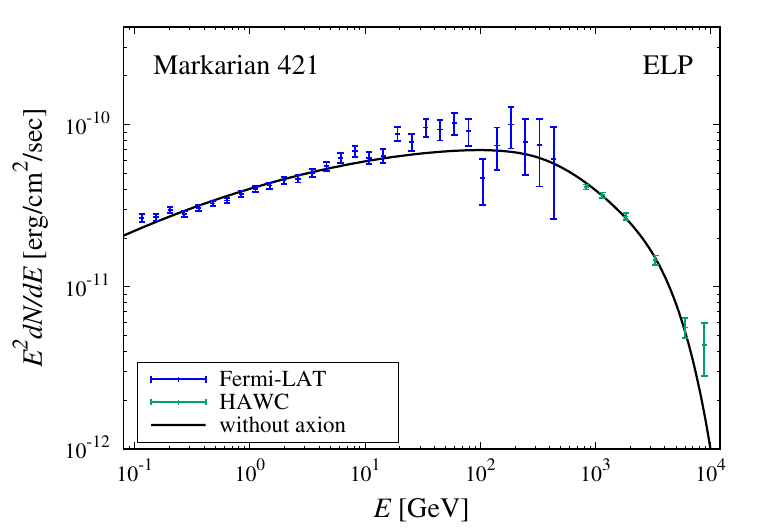}
\caption{The best-fit SEDs of Markarian\,421 under the null hypothesis with the intrinsic spectral models EPWL (left), LP (middle), and ELP (right).
The blue and green experimental data are taken from Fermi-LAT and HAWC, respectively.}
\label{fig_dNdE_mrk421_SM}
\end{figure*}

Using these intrinsic spectral models, we fit the gamma-ray of Markarian\,421 and list the corresponding best-fit chi-squares in Table~\ref{tab_chi-square_421_SM}.
Due to the smallest value of $\chi^2_{\rm null}/{\rm d.o.f.}$ for SEPWL, we finally select this spectral model to fit Markarian\,421.
The best-fit SEDs for other spectral models are also shown in Fig.~\ref{fig_dNdE_mrk421_SM}.

\begin{table}[h]
\caption{The best-fit chi-square values under the null hypothesis of Markarian\,421.}
\begin{ruledtabular}
\begin{tabular}{lccr}
Model & parameter number & $\chi^2_{\rm null}$ & $\chi^2_{\rm null}/{\rm d.o.f.}$ \\
\hline
EPWL     &   3    & 158.97   &  4.82  \\
SEPWL   &  4    &    44.24   &  1.38  \\
LP           &  4    &  244.52   &  7.64  \\
ELP         &  5    &   63.25    &  2.04  \\
\end{tabular}
\end{ruledtabular}
\label{tab_chi-square_421_SM}
\end{table}

\section{Final photon survival probability}
\label{sm_sec2}

\begin{figure*}[h]
\centering
\includegraphics[width=0.333\textwidth]{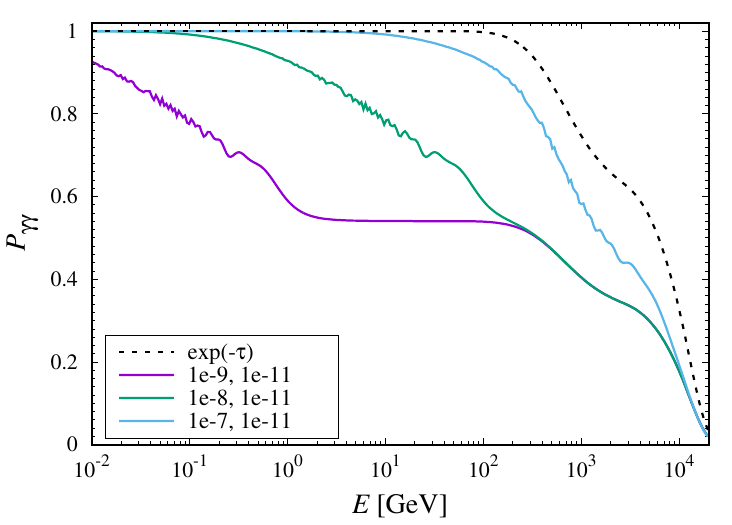}\includegraphics[width=0.333\textwidth]{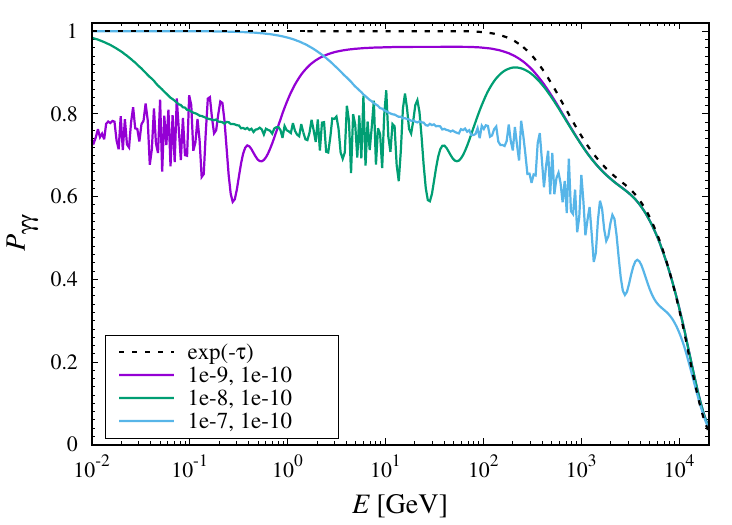}\includegraphics[width=0.333\textwidth]{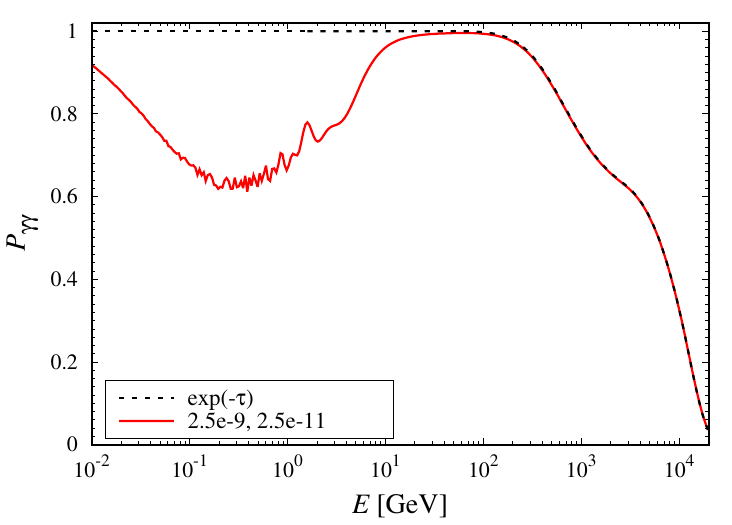}
\caption{The final photon survival probability for Markarian\,421.
The dashed line represents the pure EBL absorption effect with a factor $e^{-\tau}$.
The solid lines correspond to the final photon survival probability with typical examples.
The label $\{x, \, y\}$ represents the parameter set $\{m_a = x \, {\rm eV}, \, g_{a\gamma} = y \, \rm GeV^{-1}\}$.
Note that the right panel corresponds to $\chi^2_{\rm min}$.}
\label{fig_pa_421_SM}
\end{figure*}

In the transverse homogeneous magnetic field $B_T$, the axion-photon conversion probability can be described by $\mathcal{P}_{a\gamma}=(g_{a\gamma}B_T L_{\rm osc}/2\pi)^2 \sin^2(\pi x_3/L_{\rm osc})$ with the oscillation length
\begin{eqnarray}
L_{\rm osc} = 2\pi\left[\left[ \dfrac{|m_a^2-\omega_{\rm pl}^2|}{2E}+E\left(\dfrac{7\alpha}{90\pi} \left( \dfrac{B_T}{B_{\rm cr}}\right)^2 +\rho_{\rm CMB}\right)\right]^2+g_{a\gamma}^2B_T^2\right]^{-1/2}\, ,
\end{eqnarray}  
where $\omega_{\rm pl}$ is the plasma frequency, $E$ is the axion/photon energy, $\alpha$ is the fine-structure constant, $B_{\rm cr}$ is the critical magnetic field, and $\rho_{\rm CMB}$ represents the cosmic microwave background (CMB) photon dispersion effect. 
Here we show in Fig.~\ref{fig_pa_421_SM} the final photon survival probability $\mathcal{P}_{\gamma\gamma}$ for Markarian\,421.
In the left and middle panels, we show $\mathcal{P}_{\gamma\gamma}$ for several typical axion-photon coupling $\{m_a, \, g_{a\gamma}\}$ parameter sets.
In addition, we also show in the right panel the final photon survival probability for the minimum best-fit chi-square $\chi^2_{\rm min}=31.47$ in the $\{m_a, \, g_{a\gamma}\}$ plane, with $\{m_a \simeq 2.5\times10^{-9} \, {\rm eV}, \, g_{a\gamma} \simeq 2.5\times 10^{-11} \rm \, GeV^{-1}\}$.

\section{TS distribution}
\label{sm_sec3}

\begin{figure*}[h]
\centering
\includegraphics[width=0.44\textwidth]{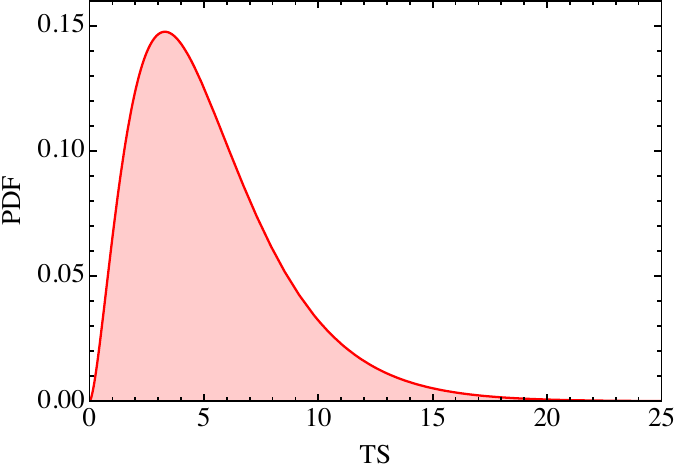}\quad\includegraphics[width=0.43\textwidth]{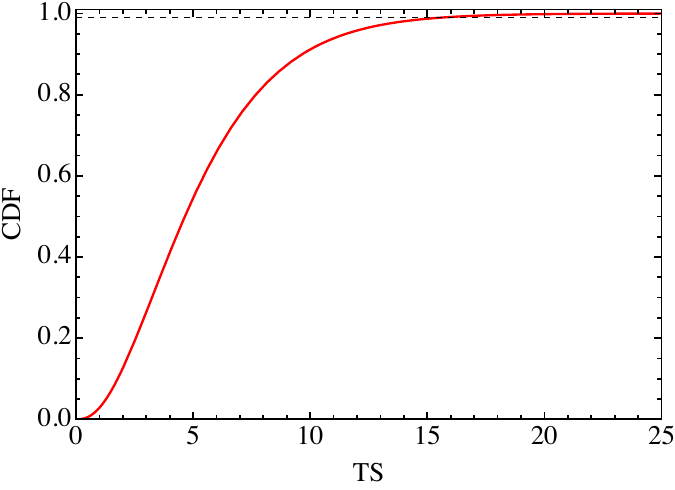}
\caption{The TS distribution with the PDF (left) and CDF (right).
The dashed line corresponds to $\rm CDF = 0.99$.}
\label{fig_TS_SM}
\end{figure*}

We also show the test statistic (TS) distribution in Fig.~\ref{fig_TS_SM} with the probability density function (PDF) and cumulative distribution function (CDF).
In our simulations, we have the non-centrality $\lambda=0.01$ and the effective $\rm d.o.f. = 5.29$, indicating the 99\% $\rm C.L.$ chi-square difference $\Delta\chi^2_{99\%}\simeq15.51$, and the threshold chi-square $\chi^2_{99\%}=31.47+15.51=46.98$.
Then we use this value to set the 99\% $\rm C.L.$ limit in the axion parameter plane.

\section{Impact of the parameter uncertainty of $B_0$}
\label{sm_sec4}

\begin{figure*}[h]
\centering
\includegraphics[width=0.49\textwidth]{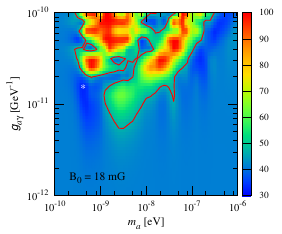}\includegraphics[width=0.49\textwidth]{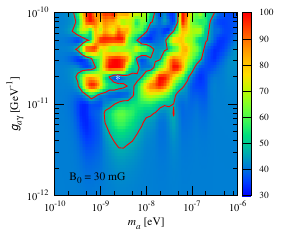}
\caption{The best-fit $\chi_{\rm axion}^2$ distribution in the $\{m_a, \, g_{a\gamma}\}$ plane for Markarian\,421 with different $B_0$.
Left: we set $B_0=18 \, \rm mG$, the label ``$*$" corresponds to $\chi^2_{\rm min}=32.94$ at $\{m_a \simeq 4.0\times10^{-10} \, {\rm eV}, \, g_{a\gamma} \simeq 1.6\times 10^{-11} \rm \, GeV^{-1}\}$. 
Right: we set $B_0=30 \, \rm mG$, the label ``$*$" corresponds to $\chi^2_{\rm min}=31.87$ at $\{m_a \simeq 2.5\times10^{-9} \, {\rm eV}, \, g_{a\gamma} \simeq 2.0\times 10^{-11} \rm \, GeV^{-1}\}$.}
\label{fig_contour_mrk421_SM}
\end{figure*}

Here we show the impact of the uncertainty of $B_0$. 
Ws show in Fig.~\ref{fig_contour_mrk421_SM} the best-fit chi-square $\chi_{\rm axion}^2$ distribution in the $\{m_a, \, g_{a\gamma}\}$ plane with different values of $B_0$.
 
\section{Axion analysis for Markarian\,501}
\label{sm_sec5}
 
\begin{figure*}[h]
\centering
\includegraphics[width=0.49\textwidth]{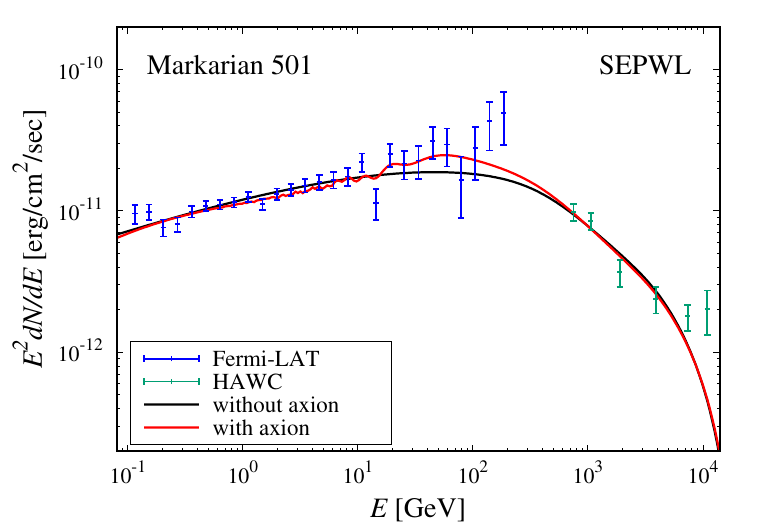} 
\caption{The best-fit SEDs of Markarian\,501 under the null and axion hypotheses.
The black and red lines represent the best-fit SEDs without/with the axion-photon conversion, respectively.
The blue and green experimental data are taken from Fermi-LAT and HAWC, respectively.}
\label{fig_dNdE_mrk501_SM}
\end{figure*}

Additionally, since the VHE gamma-ray observations of another TeV blazar Markarian\,501 (Mrk\,501, with redshift $z=0.034$) at the same time are also measured by Fermi-LAT and HAWC, in the following we make the axion analysis for this source.
The experimental data of Fermi-LAT (27 points) and HAWC (6 points) are shown in Fig.~\ref{fig_dNdE_mrk501_SM}.
Here we also select the gamma-ray intrinsic spectral model SEPWL to fit Markarian\,501, and the best-fit SED corresponds to $\chi^2_{\rm null}$ is shown in Fig.~\ref{fig_dNdE_mrk501_SM}.
The best-fit chi-squares for other spectral models are listed in Table~\ref{tab_chi-square_501_SM}.

\begin{figure*}[h]
\centering
\includegraphics[width=0.49\textwidth]{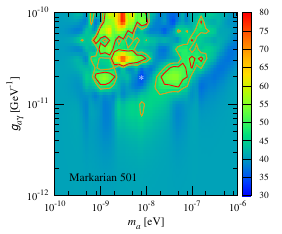}
\caption{The best-fit $\chi_{\rm axion}^2$ distribution in the $\{m_a, \, g_{a\gamma}\}$ plane for Markarian\,501. 
The red and orange contours represent the 99\% and 95\% $\rm C.L.$ exclusion regions, respectively.
The label ``$*$" corresponds to $\chi^2_{\rm min}=33.56$ at $\{m_a \simeq 7.9\times10^{-9} \, {\rm eV}, \, g_{a\gamma} \simeq 2.0\times 10^{-11} \rm \, GeV^{-1}\}$.}
\label{fig_contour_mrk501_SM}
\end{figure*}

For the blazar jet magnetic field of Markarian\,501, we take $B_0=20\pm5 \, \rm mG$, $n_0\simeq1\times10^3\, \rm cm^{-3}$, $R_{\rm VHE}=1.0\times10^{17}\, \rm cm$, $\theta_{\rm jet}=3.0^\circ$, $r_{\rm VHE}\simeq19.1\times10^{17}\, \rm cm$, and $\delta_{\rm D}=13\pm0.7$.
Then we obtain the best-fit chi-square $\chi_{\rm axion}^2$ distribution in the $\{m_a, \, g_{a\gamma}\}$ plane, which is shown in Fig.~\ref{fig_contour_mrk501_SM}.
The value of the minimum best-fit chi-square in the $\{m_a, \, g_{a\gamma}\}$ plane is $\chi^2_{\rm min}=33.56$ at $\{m_a \simeq 7.9\times10^{-9} \, {\rm eV}, \, g_{a\gamma} \simeq 2.0\times 10^{-11} \rm \, GeV^{-1}\}$, see also the corresponding SED in Fig.~\ref{fig_dNdE_mrk501_SM}.
In our TS analysis, we have the non-centrality $\lambda=0.01$ and the effective $\rm d.o.f. = 6.11$, indicating the 99\% $\rm C.L.$ threshold chi-square $\chi^2_{99\%}=50.60$, corresponding to the red contour in Fig.~\ref{fig_contour_mrk501_SM}.
Since it shows a weak limit, we also show the 95\% $\rm C.L.$ limit ($\chi^2_{95\%}=46.38$) with the orange contour.
While for Markarian\,421, since the value of $\chi^2_{95\%}$ is smaller than the best-fit chi-square under the null hypothesis $\chi^2_{\rm null}$, we just show the 99\% $\rm C.L.$ limit.
In summary, here we do not obtain a stringent upper limit from the source Markarian\,501.

\begin{table}[h]
\caption{The best-fit chi-square values under the null hypothesis of Markarian\,501.}
\begin{ruledtabular}
\begin{tabular}{lccr}
Model & parameter number & $\chi^2_{\rm null}$ & $\chi^2_{\rm null}/{\rm d.o.f.}$ \\
\hline
EPWL     &   3    &  74.91   &   2.50   \\
SEPWL   &  4    &   42.06   &  1.45    \\
LP           &  4    &   53.61   &  1.85    \\
ELP         &  5    &   50.68   &  1.81    \\
\end{tabular}
\end{ruledtabular}
\label{tab_chi-square_501_SM}
\end{table}

\section{Experimental gamma-ray data}
\label{sm_sec6}

Finally, we attach the experimental gamma-ray data of Markarian\,421 and Markarian\,501 measured by Fermi-LAT and HAWC in Tables~\ref{tab_data_421_SM} and \ref{tab_data_501_SM}, respectively.

\begin{table}[h]
\caption{The experimental gamma-ray data of Markarian\,421. The third and fourth columns are the flux and its uncertainty.}
\begin{ruledtabular}
\begin{tabular}{lccr}
Collaboration & $E$ (GeV) & $E^2dN/dE$ ($\rm erg/cm^2/sec$) & $\delta$ ($\rm erg/cm^2/sec$) \\
\hline
Fermi-LAT  &  0.1152  &  2.662e-11 &  1.457e-12\\    
Fermi-LAT  &  0.1531  &  2.686e-11 &  1.297e-12\\
Fermi-LAT  &  0.2034  &  2.984e-11 &  1.220e-12\\
Fermi-LAT  &  0.2701  &  2.805e-11 &  1.186e-12\\
Fermi-LAT  &  0.3588  &  3.059e-11 &  1.204e-12\\
Fermi-LAT  &  0.4766  &  3.268e-11 &  1.247e-12\\
Fermi-LAT  &  0.633   &  3.405e-11 &  1.324e-12\\
Fermi-LAT  &  0.8409  &  3.733e-11 &  1.472e-12\\
Fermi-LAT  &  1.117   &  4.030e-11 &  1.667e-12\\
Fermi-LAT  &  1.484   &  4.214e-11 &  1.852e-12\\
Fermi-LAT  &  1.971   &  4.542e-11 &  2.143e-12\\
Fermi-LAT  &  2.618   &  4.647e-11 &  2.466e-12\\
Fermi-LAT  &  3.477   &  5.040e-11 &  2.922e-12\\
Fermi-LAT  &  4.619   &  5.540e-11 &  3.508e-12\\
Fermi-LAT  &  6.135   &  6.246e-11 &  4.246e-12\\
Fermi-LAT  &  8.15    &  6.849e-11 &  5.077e-12\\
Fermi-LAT  &  10.82   &  6.250e-11 &  5.583e-12\\
Fermi-LAT  &  14.38   &  6.444e-11 &  6.508e-12\\
Fermi-LAT  &  19.1    &  8.767e-11 &  8.731e-12\\
Fermi-LAT  &  25.37   &  7.798e-11 &  9.440e-12\\
Fermi-LAT  &  33.7    &  9.612e-11 &  1.198e-11\\
Fermi-LAT  &  44.76   &  9.335e-11 &  1.353e-11\\
Fermi-LAT  &  59.46   &  1.023e-10 &  1.630e-11\\
Fermi-LAT  &  78.98   &  9.133e-11 &  1.762e-11\\
Fermi-LAT  &  104.9   &  4.667e-11 &  1.477e-11\\
Fermi-LAT  &  139.4   &  7.443e-11 &  2.167e-11\\
Fermi-LAT  &  185.1   &  1.001e-10 &  2.888e-11\\
Fermi-LAT  &  245.9   &  7.840e-11 &  2.949e-11\\
Fermi-LAT  &  326.6   &  7.500e-11 &  3.342e-11\\
Fermi-LAT  &  433.8   &  6.148e-11 &  3.538e-11\\
HAWC       &  831.3   &  4.124e-11 &  1.474e-12\\
HAWC       &  1145    &  3.678e-11 &  1.371e-12\\
HAWC       &  1835    &  2.713e-11 &  1.323e-12\\
HAWC       &  3306    &  1.463e-11 &  9.799e-13\\
HAWC       &  5984    &  5.615e-12 &  7.881e-13\\
HAWC       &  8809    &  4.379e-12 &  1.578e-12\\    
\end{tabular}
\end{ruledtabular}
\label{tab_data_421_SM}
\end{table}

\begin{table}[h]
\caption{Same as Table~\ref{tab_data_421_SM} but for Markarian\,501.}
\begin{ruledtabular}
\begin{tabular}{lccr}
Collaboration & $E$ (GeV) & $E^2dN/dE$ ($\rm erg/cm^2/sec$) & $\delta$ ($\rm erg/cm^2/sec$) \\
\hline
Fermi-LAT & 0.1153 & 9.598e-12 & 1.459e-12\\ 
Fermi-LAT & 0.1531 & 9.851e-12 & 1.259e-12\\
Fermi-LAT & 0.2034 & 7.649e-12 & 1.041e-12\\
Fermi-LAT & 0.2701 & 8.056e-12 & 9.441e-13\\
Fermi-LAT & 0.3588 & 9.912e-12 & 9.147e-13\\
Fermi-LAT & 0.4764 & 1.083e-11 & 8.899e-13\\
Fermi-LAT & 0.6332 & 1.112e-11 & 8.724e-13\\
Fermi-LAT & 0.8408 & 1.152e-11 & 9.082e-13\\
Fermi-LAT & 1.117  & 1.256e-11 & 9.967e-13\\
Fermi-LAT & 1.484  & 1.117e-11 & 1.039e-12\\
Fermi-LAT & 1.971  & 1.319e-11 & 1.219e-12\\
Fermi-LAT & 2.618  & 1.413e-11 & 1.436e-12\\
Fermi-LAT & 3.477  & 1.513e-11 & 1.609e-12\\
Fermi-LAT & 4.62   & 1.604e-11 & 1.935e-12\\
Fermi-LAT & 6.137  & 1.662e-11 & 2.205e-12\\
Fermi-LAT & 8.152  & 1.755e-11 & 2.624e-12\\
Fermi-LAT & 10.82  & 2.217e-11 & 3.312e-12\\
Fermi-LAT & 14.38  & 1.144e-11 & 2.829e-12\\
Fermi-LAT & 19.1   & 2.516e-11 & 4.681e-12\\
Fermi-LAT & 25.37  & 2.165e-11 & 4.962e-12\\
Fermi-LAT & 33.7   & 2.281e-11 & 5.872e-12\\
Fermi-LAT & 44.75  & 3.127e-11 & 7.846e-12\\
Fermi-LAT & 59.47  & 2.949e-11 & 8.848e-12\\
Fermi-LAT & 78.99  & 1.660e-11 & 7.688e-12\\
Fermi-LAT & 104.9  & 2.801e-11 & 1.143e-11\\
Fermi-LAT & 139.4  & 4.310e-11 & 1.626e-11\\
Fermi-LAT & 185.1  & 4.946e-11 & 2.010e-11\\
HAWC       & 749.8  & 9.850e-12 & 1.318e-12\\
HAWC       & 1060   & 8.533e-12 & 1.155e-12\\
HAWC       & 1890   & 3.696e-12 & 8.000e-13\\
HAWC       & 3890   & 2.397e-12 & 5.134e-13\\
HAWC       & 7362   & 1.797e-12 & 3.769e-13\\
HAWC       & 10900  & 2.035e-12 & 7.075e-13\\   
\end{tabular}
\end{ruledtabular}
\label{tab_data_501_SM}
\end{table} 
 
\end{document}